\documentclass{ws-ijmpa}

\begin{document}

\markboth{R. Kami\'nski {\em et al.}}
{New dispersion relations in the description of $\pi\pi$ scattering amplitudes}

%
\catchline{}{}{}{}{}
%

\title{NEW DISPERSION RELATIONS IN THE DESCRIPTION OF $\pi\pi$ SCATTERING AMPLITUDES}

\author{\footnotesize R. Kami\'nski$^{a}$$^,$\footnote{E-mail address: 
Robert.Kaminski@ifj.edu.pl}~, 
R. Garc\'\i{}a-Mart\'\i{}n$^{b}$,
P. Grynkiewicz$^{a}$
J. R. Pel\'aez$^{b}$, 
F. J. Yndur\'ain$^{c}$ 
}

\address{
$^{a}$Department of Theoretical Physics, H. Niewodnicza\'nski Institute
 of Nuclear Physics, Polish Academy of Sciences, 31-342 
 Krak\'ow, Poland\\ 
$^{b}$Departamento de F\'{\i}sica Te\'orica~II
 (M\'etodos Matem\'aticos),
Facultad de Ciencias F\'{\i}sicas,
Universidad Complutense de Madrid,
E-28040, Madrid, Spain\\
$^{c}$Departamento de F\'{\i}sica Te\'orica, C-XI
 Universidad Aut\'onoma de Madrid,
 Canto Blanco,
E-28049, Madrid, Spain\\
}

\maketitle


\begin{abstract}
We present a set of once subtracted 
dispersion relations  which implement crossing symmetry
conditions for the $\pi\pi$ scattering amplitudes below 1~GeV.
We compare and discuss
the results obtained for the once and twice subtracted
dispersion relations, known as Roy's equations, for three 
$\pi\pi$ partial $JI$
waves, $S0$, $P$ and $S2$.
We also show
that once subtracted dispersion relations provide a stringent
test of crossing  and analyticity for 
$\pi\pi$ partial wave amplitudes, remarkably precise in the
$400$ to 1.1 GeV region, where the resulting uncertainties are significantly
smaller than those coming from standard  Roy's equations,
given the same input.
\end{abstract}

\ccode{PACS numbers: 13.75.Lb, 11.55.-m, 11.55.Fv, 11.80.Et}

\section{Introduction}  

In 1971 S. M. Roy~\cite{Roy71} derived a 
set of coupled integral equations, the Roy Equations (RE),
for the $\pi\pi$ scattering partial waves, by
implementing crossing symmetry conditions into 
twice subtracted dispersion relations.
In recent years, Roy's equations have been used in several ways:
to obtain predictions 
for low energy $\pi\pi$ scattering using Chiral Perturbation Theory (ChPT)
\cite{A4,CGLNPB01}, to test those predictions
(ChPT)~\cite{DescotesGenon:2001tn,KJYII,KJYI}, and also to
eliminate the well known "up-down"
ambiguity~\cite{Pennington1973ab,KLLUpDown}.

In a series of works, our group~\cite{KJYII,KJYI} has also used a dispersive approach,
to obtain, using also the most recent experimental results,
 a precise data parametrization of $\pi\pi$ scattering amplitudes 
consistent with analyticity, unitarity and crossing.
In fact, the recent data from E865 collaboration 
at Brookhaven~\cite{Pislak2001} and from NA48/2~\cite{NA482008} 
provide us with new and very precise information on the $\pi\pi$ scattering at low energies.
In our works we have combined Forward Dispersion Relations (FDR) and Roy's Eqs.
Let us remark that we have only used the very general 
properties of analyticity, crossing, etc... and data, 
so that the approach is model independent. Furthermore, we have not 
included ChPT constraints, so that our results  could be used as
tests of ChPT. The
advantage of FDR is that 
they are very precise, can be extended up to any energy 
and do not depend on the large $t$ behavior. In contrast, Roy Eqs. use 
the full $t$ dependence since they are written in terms of partial waves
and can only be used up to roughly 1.1 GeV. However, RE provide a
simple and well defined analytic extension of partial waves 
for the calculation of poles in the complex plane.
Such analytic extension of $\pi\pi$ scattering partial waves 
to the complex plane is of particular interest 
for the understanding of
the controversial sigma resonance. 
Actually, Roy's equations have been used to predict
very precisely the sigma pole position~\cite{Caprini}
using the ChPT determination of the scattering lengths.

We report here about our work in progress to improve our description
of the energy region above $400$~MeV, that can subsequently
provide a precise determination of the sigma pole.
Actually, when using standard RE, 
the large experimental error of the scattering length $a^2_0$ 
of the isospin 2 scalar partial wave, becomes a very
large uncertainty in the intermediate energy region and
for the sigma pole determination. 
For this reason we briefly describe here a new set of
once-subtracted RE, 
denoted GKPY Eqs. for brevity, and we show the relative
sizes of the different contributions, comparing them with those for standard RE.
We show that, given the same input,
the uncertainties of standard Roy's Eqs. are smaller than those of GKPY Eqs.
at low energies. However, the uncertainties of the once-subtracted GKPY Eqs. are
smaller than those of Roy's Eqs. above, roughly 400~MeV, up to 1.1~GeV.
Hence, in that energy region, GKPY provide a very precise 
additional constraint for our dispersive analysis
of data, and a very precise analytic extension to determine the position
of the sigma pole from experiment.

\section{Once and Twice Subtracted Dispersion Relations}

A twice subtracted dispersion relation for the scattering amplitude $T(s,t)$
of a given process
is an expression of the form:
\begin{eqnarray}
  \nonumber
   Re T(s) & = & g(s_1, s_2) + h(s;s_1, s_2) 
+   \frac{(s_1-s)(s_2-s)}{\pi}
   \int_{s_{th}}^\infty
   \frac{Im T(s\prime)}{(s\prime - s)(s\prime-s_1)(s\prime-s_2)} ds\prime \\
   & + &
   \frac{(s_1-s)(s_2-s)}{\pi}
   \int_{-t}^{-\infty}
   \frac{Im T(s\prime)}{(s\prime - s)(s\prime-s_1)(s\prime-s_2)} ds\prime
\label{2S}
\end{eqnarray}
which relates the real part of an amplitude for a real $s$ value to the imaginary
part of the amplitude integrated over the whole energy range, together with two functions,
$g(t; s_1, s_2)$ and $h(s, t; s_1, s_2)$, called the subtraction terms, $ST(s)$.
In the paper by Roy~\cite{Roy71} such a relation is written for the definite
$\pi\pi$ scattering isospin amplitudes, in a slightly modified way
to show explicitly the crossing relations between the $s$ and $u$
channels, and in which the subtraction points are taken to be $s_1=s_2=0$.
In addition, and for convenience,
the three isospin amplitudes are written as an isospin vector
amplitude $\vec T(s,t)=(T^0,T^1,T^2)$. This provides a relation among all the
isospin processes by means of three crossing matrices $C_{st}$, $C_{tu}$, $C_{su}$,
defined as:
\begin{equation}
  \vec T(s,t,u)=C_{st}\vec T(t,s,u)=C_{su}\vec T(u,t,s)=C_{tu}\vec T(s,u,t).
\end{equation}
By using these, $s\leftrightarrow u$ crossing symmetry and the fact that
on the $t$ channel the amplitudes with given isospin are of definite
symmetry, the subtraction constants can be rewritten as
$C_{st} [ \vec C(t) + (s-u) \vec D(t) ]$,
with $\vec C(t)=(c^0(t),0,c^1(t))$ and $\vec D(t)=(0,d(t),0)$.
Thus,
\begin{equation}
C_{st} T(s=0, t = t_0, u = 4m_{\pi}^2-t_0) = T(s=t_0, t = 0, u = 4m_{\pi}^2-t_0),
\label{CrossingSymmetry}
\end{equation}
which leads to:
\begin{eqnarray}
  \nonumber
  C_{st} T(0,t,4m_{\pi}^2-t)
  & = & 
  \frac{1}{\pi}\int_{4m_{\pi}^2}^{\infty} ds\prime \frac{C_{st}C_{su}Im \vec T(s\prime,t)}{s\prime^2}
  \frac{(4m_{\pi}^2-t)^2}{s'-4m_{\pi}^2+t} \\ \nonumber
  & + & \vec C(t) + (t-4m_\pi^2)\vec D(t),
  \\ \nonumber   
  T(t,0,4m_{\pi}^2-t)
  & = & 
  \frac{1}{\pi}\int_{4m_{\pi}^2}^{\infty} ds\prime
  \left(\frac{t^2}{s'-t} + \frac{(4m_{\pi}^2-t)^2}{s'-4m_{\pi}^2+t}C_{su}\right)
  \frac{Im \vec T(s\prime,0)}{s\prime^2} \\
  & + &
  C_{st}[\vec C(0) + (2t-4m_\pi^2)\vec D(0)].
\label{RoyDeriv}
\end{eqnarray}
In order to express $\vec C(t)$ and $\vec D(t)$ in terms of known quantities one
takes advantage of the fact that $(1\pm C_{tu})/2$ are orthogonal projectors over
the $s\leftrightarrow u$ symmetric or antisymmetric components, and evaluate the
amplitude at threshold:
\begin{equation}
  \vec T(4m_{\pi}^2,0,0) = 32\pi(a_0^0,0,a_0^2) = 
  C_{st} [ \vec C(0) + 4m_\pi^2 \vec D(0) ] + 
  \frac{1}{\pi}\int_{4m_\pi^2}^{\infty} ds\prime \frac{Im \vec T(s\prime,0)}{s\prime^2}
  \frac{16m_{\pi}^2}{s\prime-4m_{\pi}^2}.
\label{RoyThr}
\end{equation}

After projection into partial waves $T^I(s,t) = 32\pi\sum_{\ell} (2\ell+1) P_\ell(x(t)) f^I_\ell(s)$
one obtains the full expression for Roy's equations:
\begin{equation}
    \begin{array}{rcl}
    \mbox{Re } f_{\ell}^{I}(s) & = &
        a_{0}^{0}\delta_{I0} \delta_{\ell 0} + a_{0}^{2}\delta_{I2}
        \delta_{\ell 0}\\
     & + &  \displaystyle \frac{s-4{m_{\pi}}^2}{12{m_{\pi}}^2}(2a_{0}^{0}-5a_{0}^{2})\    
       (\delta_{I0}\delta_{\ell 0}+
        \frac{1}{6}\delta_{I1}\delta_{\ell 1} 
        -\ \frac{1}{2}\ \delta_{I2}\delta_{\ell 0})\\
     & + & 
        \displaystyle \sum\limits_{I'=0}^{2}
        \displaystyle \sum\limits_{\ell'=0}^{1}
     \hspace{0.25cm}-\hspace{-0.55cm}
        \displaystyle \int \limits_{4m_{\pi}^2}^{s_{max}}\ ds'
     K_{\ell \ell^\prime}^{I I^\prime}(s,s') \mbox{Im }f_{\ell'}^{I^\prime}
     (s') + d_{\ell}^{I}(s,s_{max})
\end{array}
\label{RoyEquations}
\end{equation}
where the integrals with the kernels 
$K^{II'}_{\ell\ell'}(s,s')$ 
contain the contributions of the $S0$, $P$ and $S2$ waves 
below $s_{max}$, and are called kernel terms, $KT(s)$.
The so called driving terms $d^I_\ell(s,s_{max})$ (abbreviated $DT(s)$)
describe the influence 
of these waves above $s_{max}$, and of the higher partial waves from the 
$\pi\pi$ threshold to infinity.
In our previous analysis,
the value $s_{max}^{1/2} = 1420$ MeV was chosen after studying
the experimental data on the $\pi\pi$
scattering \cite{KJYI}.
Above this energy a Regge parametrization is used.

The derivation of the GKPY equations follows this very same pattern,
but begins with a once subtracted
dispersion relation. This
leads to:
\begin{equation}
\begin{array}{rcl}
\mbox{Re }f_\ell^I(s) & = &        
\sum_{I'}C^{st}_{II'}a_0^{I'} 
+\sum_{\ell'}(2\ell'+1)\\
&\times&
\displaystyle \int_{4m_\pi^2}^{s_{max}} ds'
\Bigg\{K_{\ell\ell'}(s,s')Im f^I_{\ell'}(s')
-L_{\ell\ell'}(s,s')\sum_{I'}C^{su}_{II'}Im f^{I'}_{\ell'}(s')\\
& &+\,\sum_{I''}C^{st}_{II''}\left[M_\ell(s,s')Im f^{I''}_{\ell'}(s')-
N_\ell(s,s')\sum_{I'''}C^{su}_{I''I'''}Im f^{I'''}_{\ell'}(s')
\right]\Bigg\}\\
&+&\,\mbox{Re }f^{(\hbox{h.e.}),I}_\ell(s).
\end{array}
\label{1SEquations}
\end{equation}
In equations (\ref{RoyEquations}) and (\ref{1SEquations}) the 
imaginary parts on the right hand side correspond to the so called "input" amplitudes,
known in our case from experiment, while the real parts on the left hand
side correspond to the "output" from the dispersion relations.
The integrals with the kernels $K, L, N$ and $M$ and high energy parts
$Re f_\ell^{(h.e.)}(s)$ in Eq.~(\ref{1SEquations}) have the same meaning as
the kernel and driving terms, respectively, in Roy's equations.
Their expressions are lengthy and will be detailed in a future
publication.
Note that, as the once subtracted GKPY equations have kernel terms
that behave as $\sim 1/s^2$ at higher energies, instead of the
$\sim 1/s^3$ behavior in Roy's Eqs., the weight of the high energy
region is larger. However, as it is seen in Fig.~1 and explained
in the next section, it is
well under control, as the driving terms are still smaller than
the kernel terms.
For our purposes here it is enough to describe in detail just the
subtraction constant terms in the first line of Eq.~(\ref{1SEquations}).

\section{Numerical Results}

\begin{figure}[!h!]
\begin{center}
\begin{minipage}[h!t]{0.49\textwidth}
\centerline{\epsfysize 5.8 cm
\epsfbox{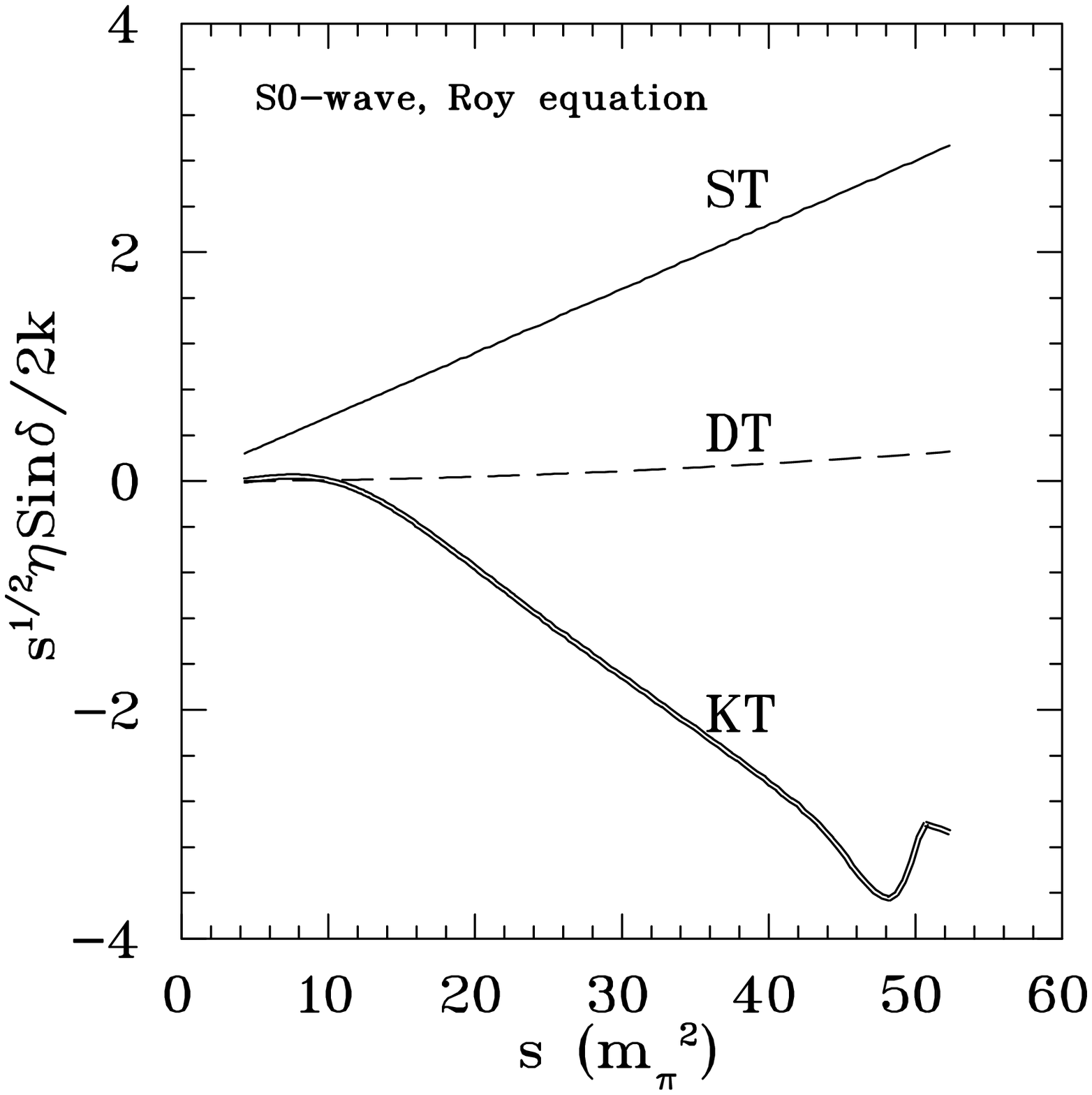}}
\end{minipage} \hfill
\begin{minipage}[h!]{0.49\textwidth}
\centerline{\epsfysize 5.8 cm
\epsfbox{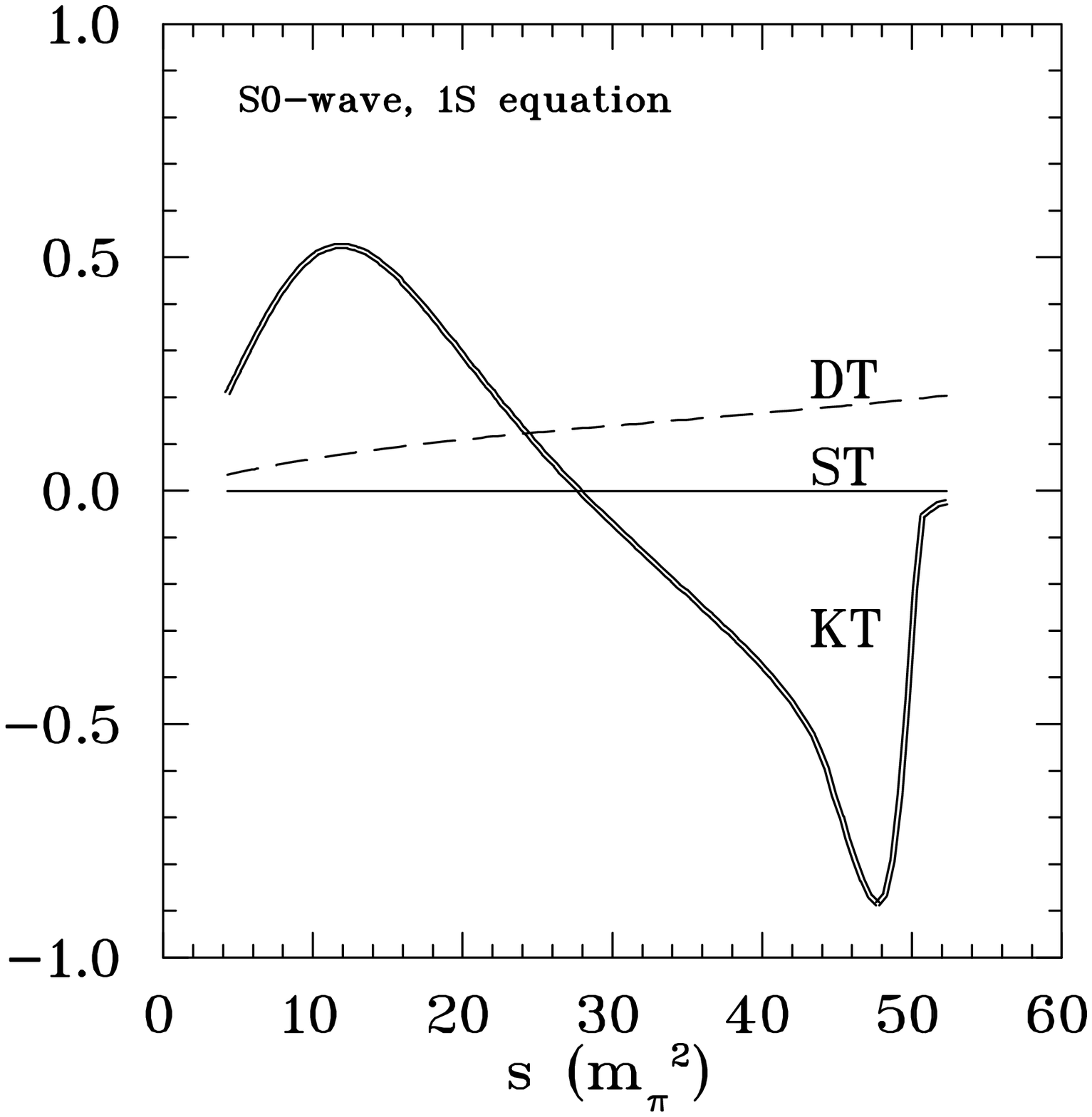}}
\end{minipage}
\begin{minipage}[h!]{0.49\textwidth}
\centerline{\epsfysize 5.8 cm
\epsfbox{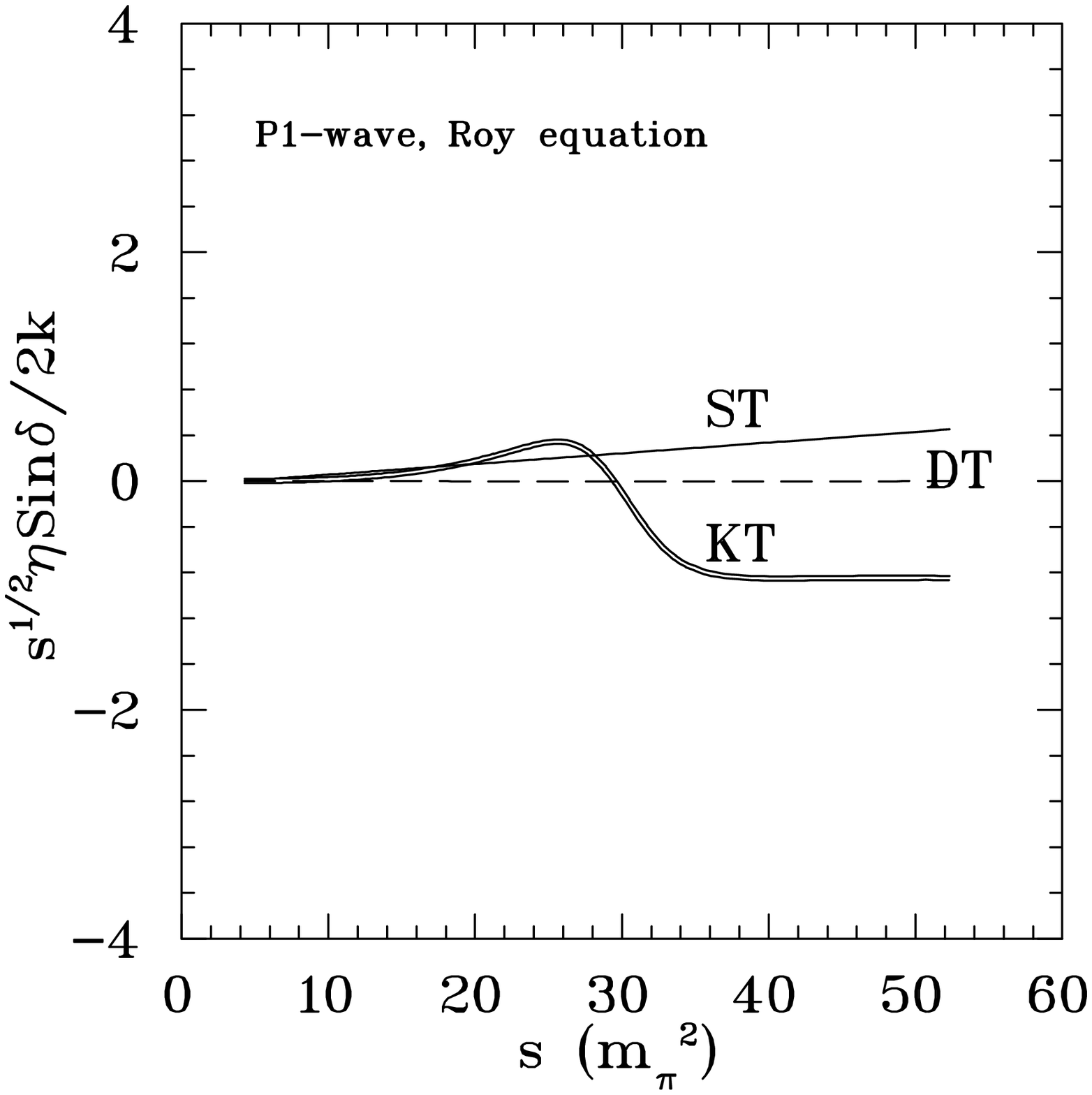}}
\end{minipage} \hfill
\begin{minipage}[h!]{0.49\textwidth}
\centerline{\epsfysize 5.8 cm
\epsfbox{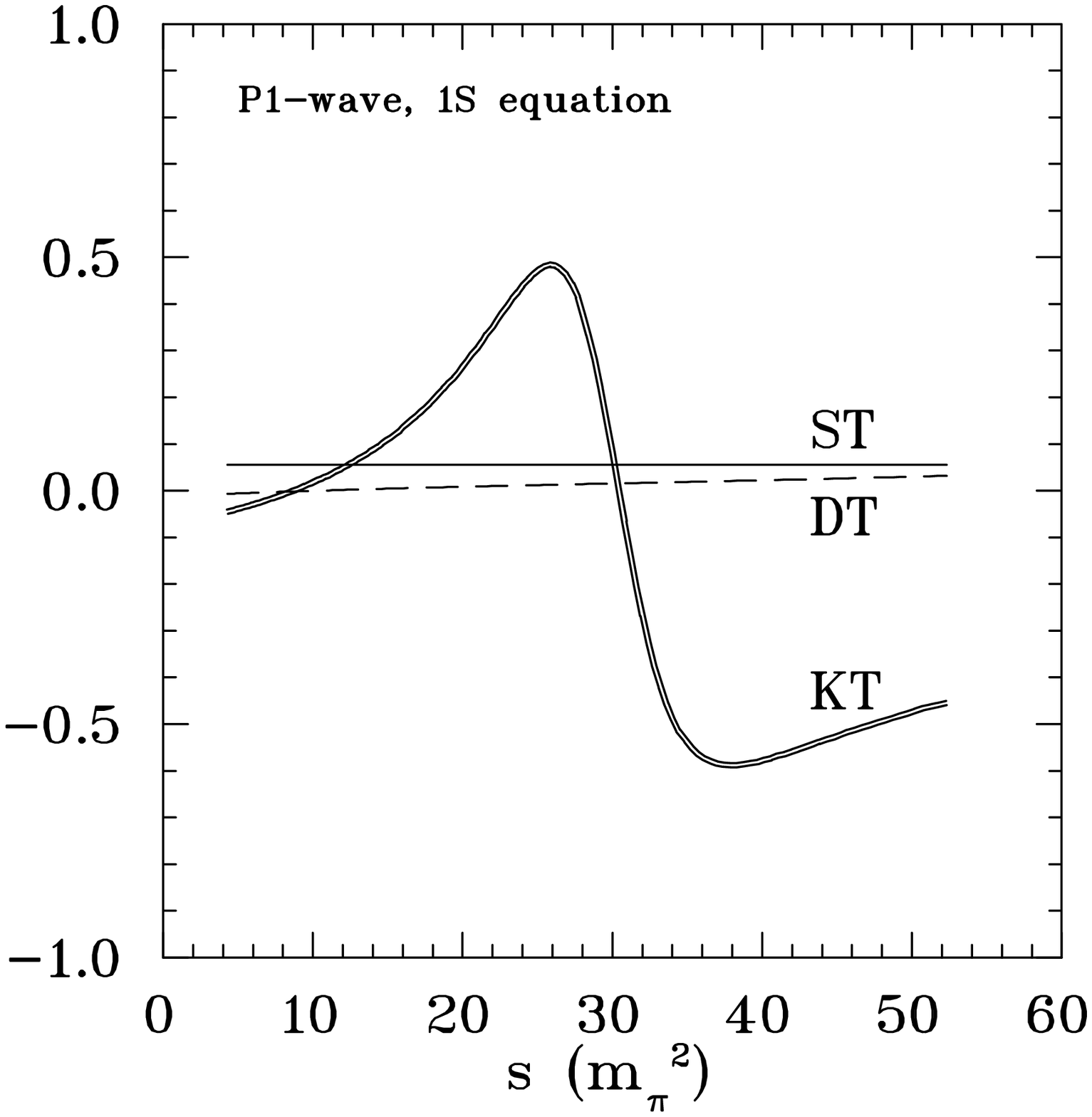}}
\end{minipage}
\begin{minipage}[h!]{0.49\textwidth}
\centerline{\epsfysize 5.8 cm
\epsfbox{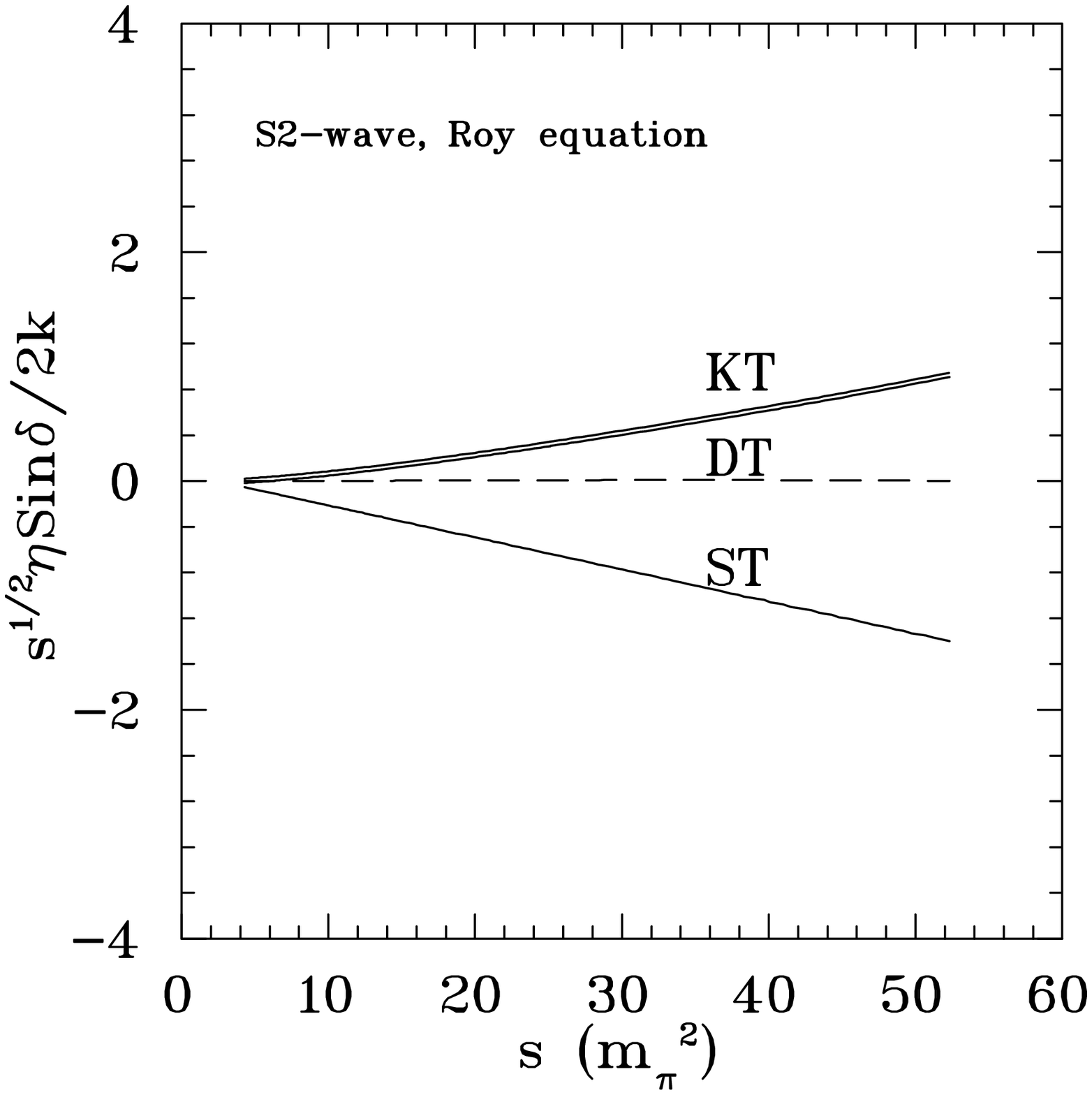}}
\end{minipage} \hfill
\begin{minipage}[h!]{0.49\textwidth}
\centerline{\epsfysize 5.8 cm
\epsfbox{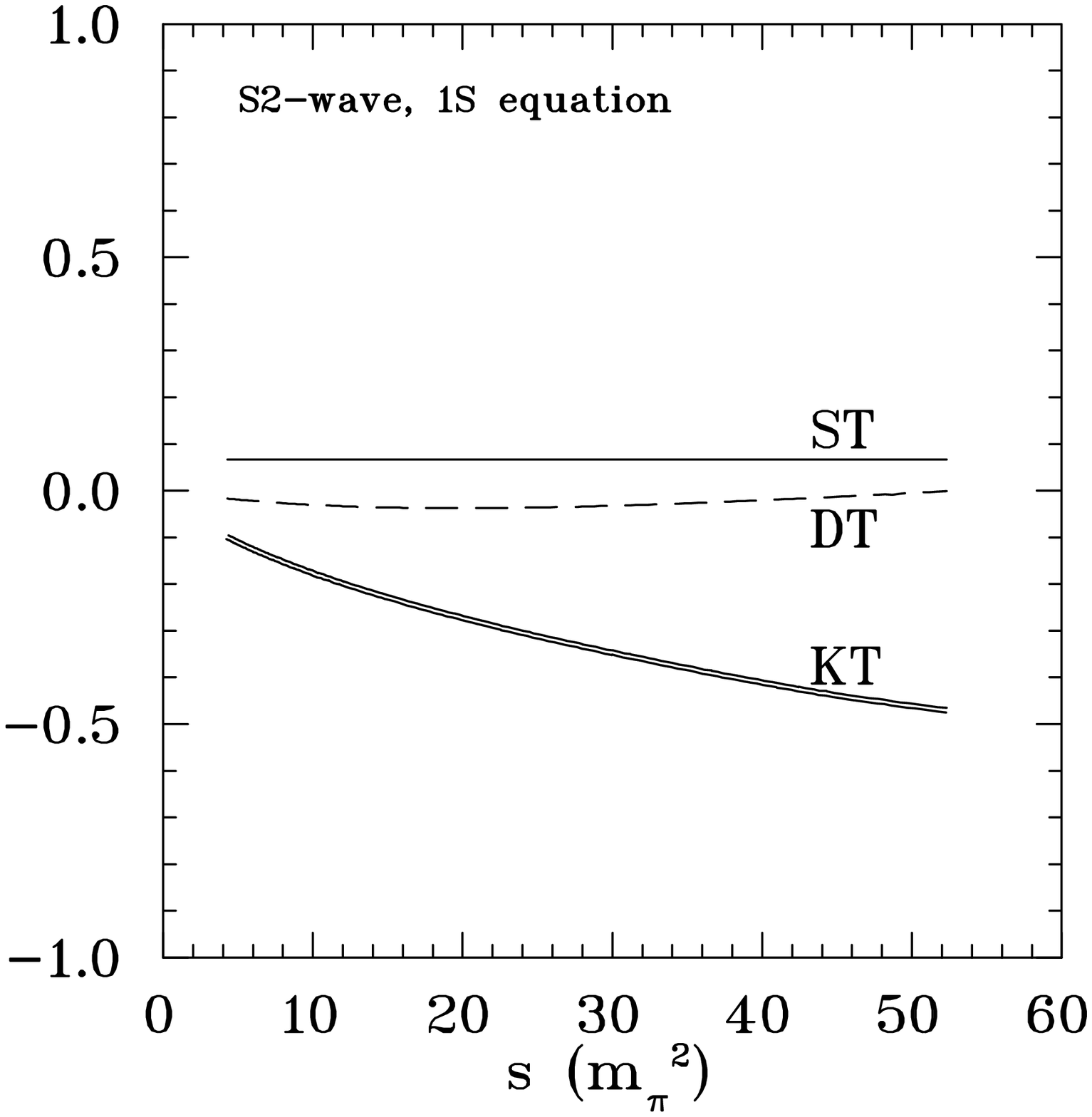}}
\end{minipage}
\caption{\label{Decompositions} 
Decomposition of the results from Roy's and 1S equations
into subtracting term ST, kernel term KT, and driving term DT for the  
$S0$-, $P$- and $S2$-waves. Note the different scales used in the left
and right columns.}
\end{center}
\end{figure}

Figure \ref{Decompositions} presents a decomposition of the equations 
(\ref{RoyEquations}) and 
(\ref{1SEquations}) into three parts: the subtracting terms $ST(s)$,
the kernel terms $KT(s)$ and the driving terms $DT(s)$.
This is done for the $S0$, $P$ and $S2$ waves. Note the different
scales on the left and right columns in the figure.
The numerical calculations have been performed by taking the
Constrained Fit to Data amplitudes fitted from experiment in~\cite{KJYII}
as input. This fit describes the experimental data well, and
has been constrained to satisfy Forward Dispersion Relations,
Roy's equations and some crossing sum rules.
As can be seen in Fig. \ref{Decompositions}, the $ST(s)$ and
$KT(s)$ terms in Roy's Eqs. become huge at higher energies and
suffer a strong cancelation. In fact, for a sufficiently large
energy, both terms are much larger than the unitarity
bound $\vert Re t\vert \le s^{1/2}/2k \sim 1$, which is only
satisfied by the real part of the total amplitude after
this strong cancelation. In the case of the $S2$ and $P$
we do not find such a huge cancelation, since both $ST(s)$
and $KT(s)$ are small enough up to energies of
about $s \approx 50 m_\pi^2 \approx (1\mbox{ GeV})^2$.

In the case of the GKPY equations for all waves, however,
the $ST(s)$ terms are constant (see eq. \ref{1SEquations}), and in fact
much smaller than the $KT(s)$ terms, which are clearly the dominant
ones. Therefore, no big cancellations between any two terms are
needed in order to reconstruct the total real part of the amplitude.
Note that, although the $DT(s)$ terms in the GKPY equations are
larger than in Roy's equations due to the fact that there is one subtraction
less, they are still small compared with the dominant term $KT(s)$.
Thus, the high energy behavior is still well under control.

\begin{figure}[h!]
\begin{center}
\begin{minipage}[h!]{0.49\textwidth}
\centerline{\epsfysize 5.7 cm
\epsfbox{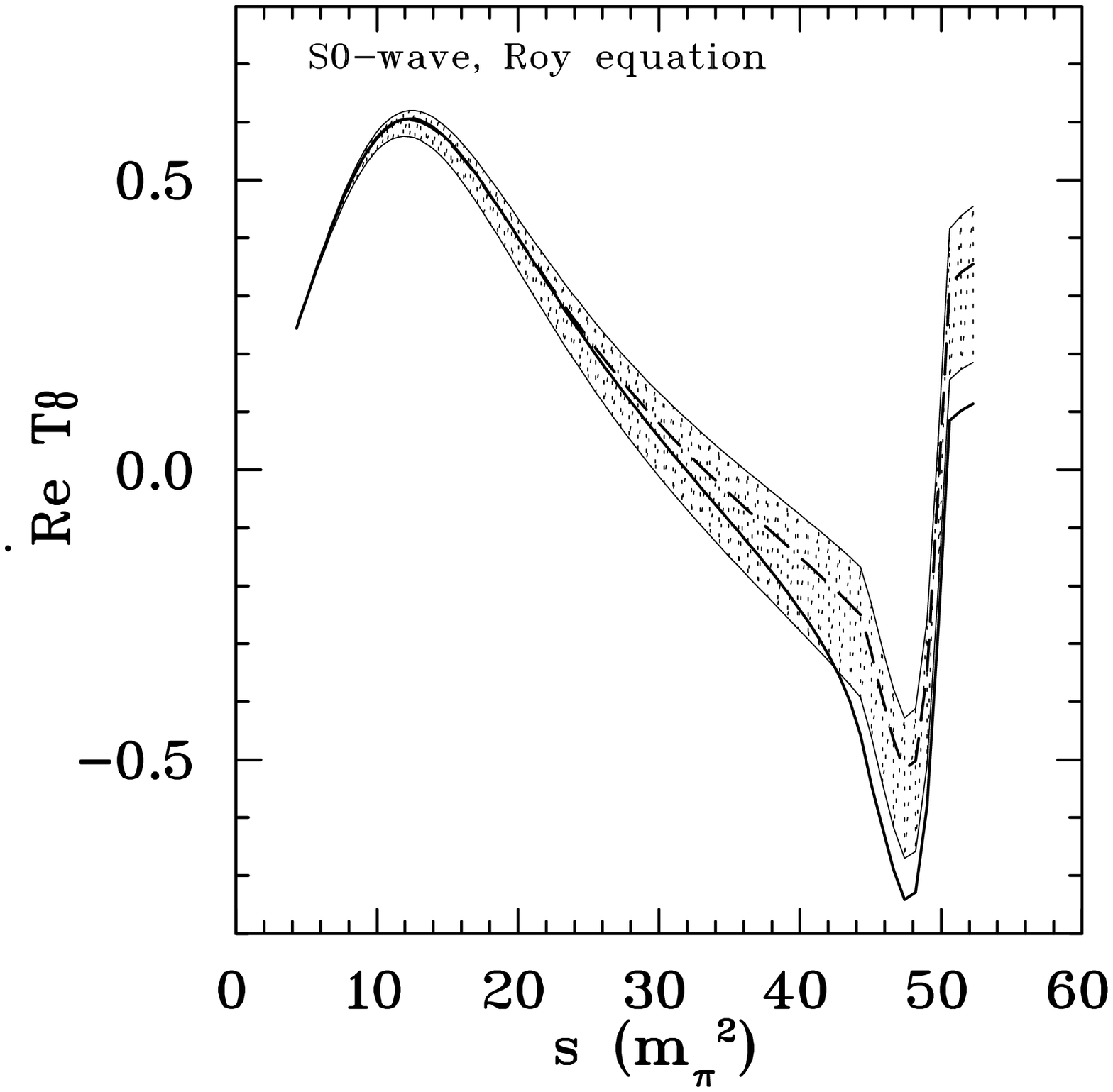}}
\end{minipage} \hfill
\begin{minipage}[h!]{0.49\textwidth}
\centerline{\epsfysize 5.7 cm
\epsfbox{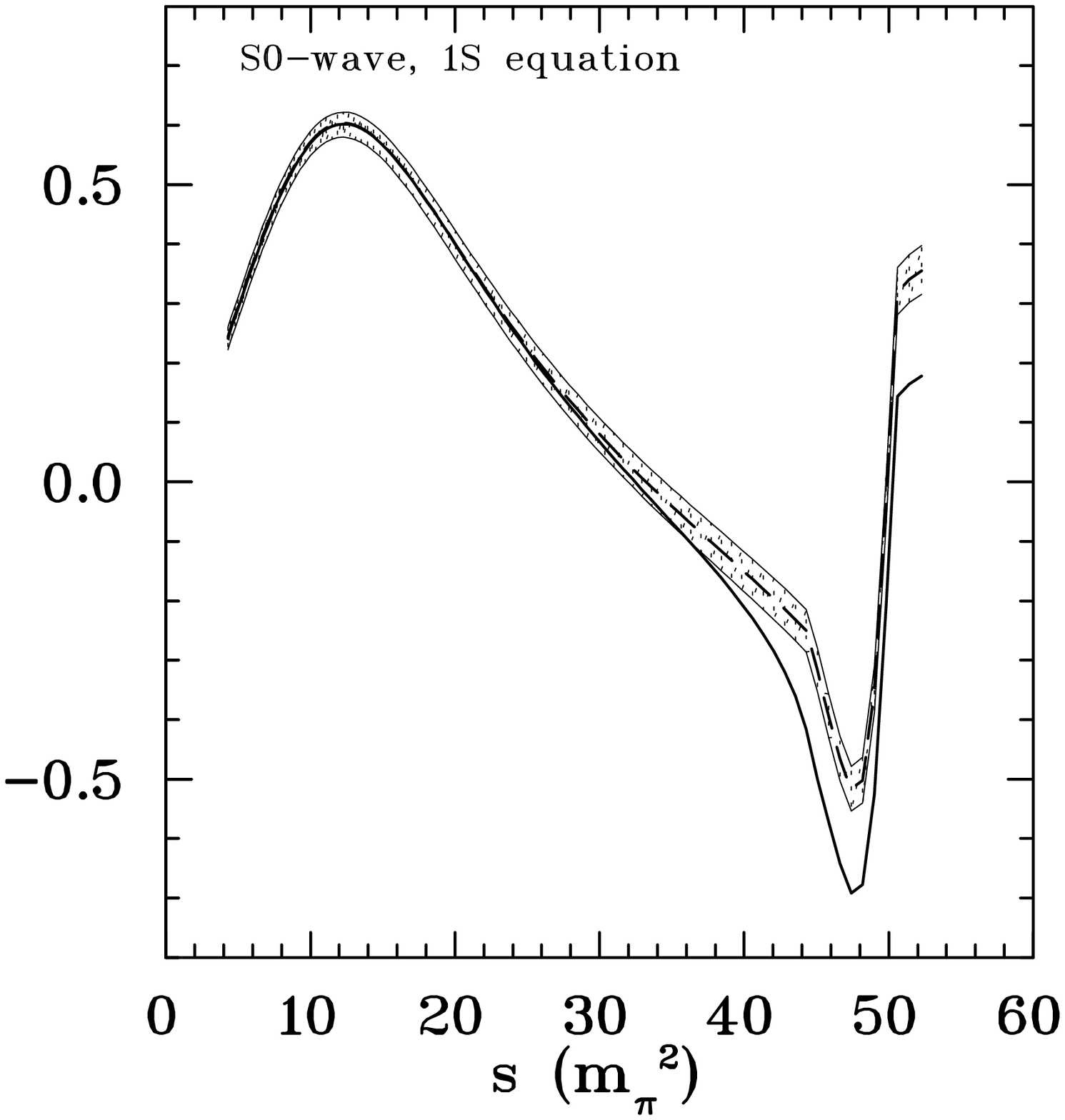}}
\end{minipage}
%


\begin{minipage}[h!]{0.49\textwidth}
\centerline{\epsfysize 5.7 cm
\epsfbox{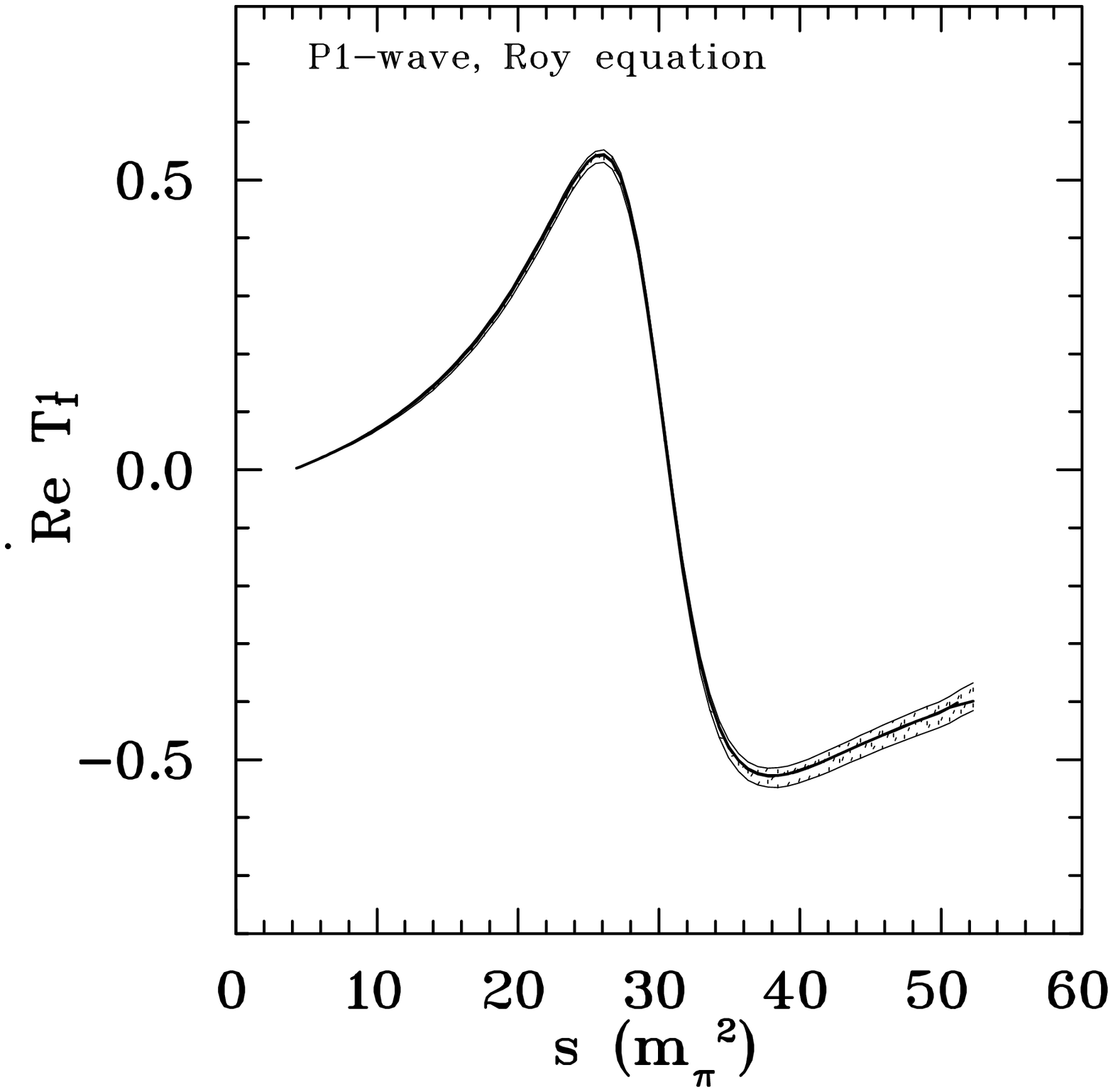}}
\end{minipage}
\begin{minipage}[h!]{0.49\textwidth}
\centerline{\epsfysize 5.7 cm
\epsfbox{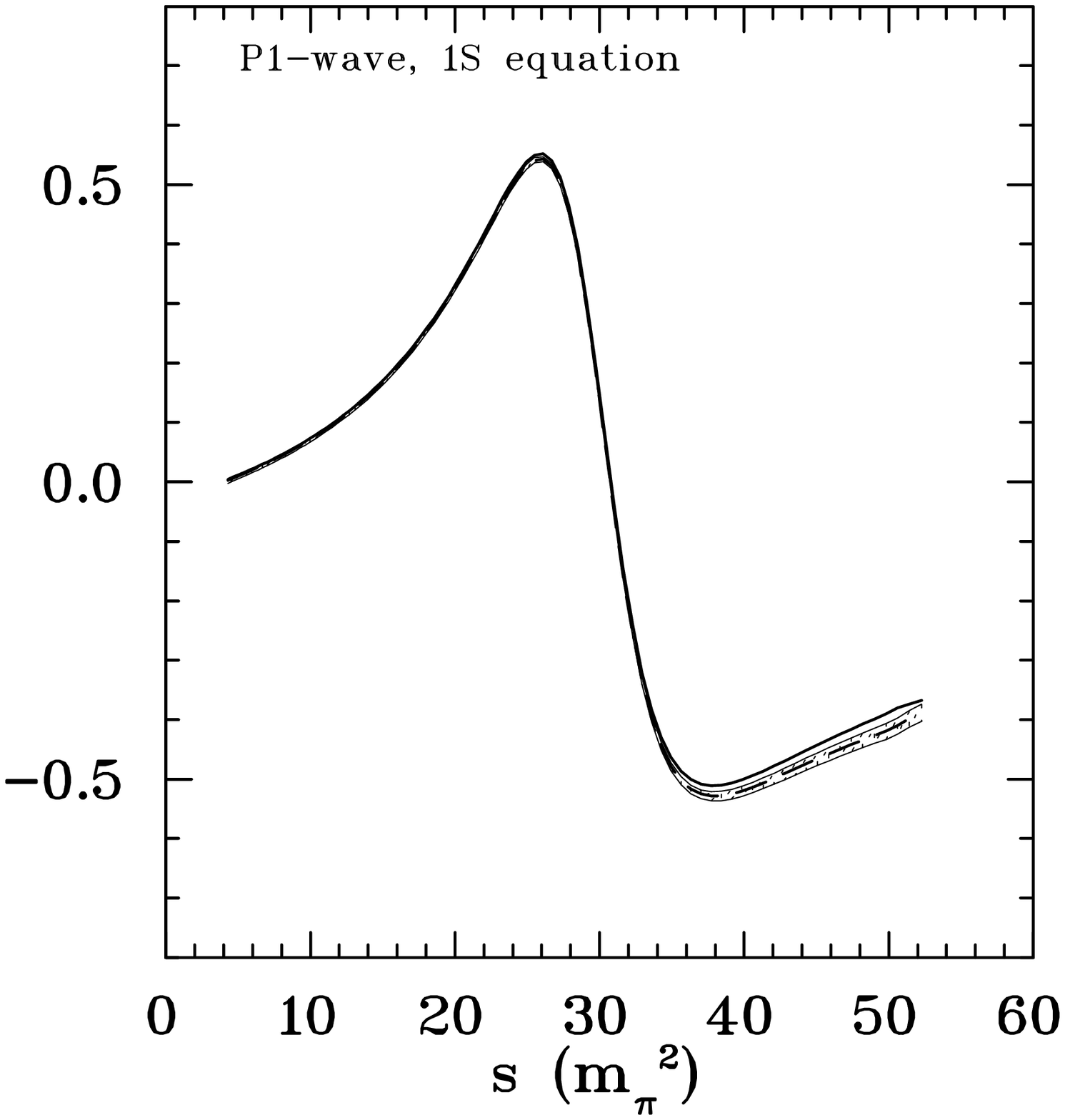}}
\end{minipage}
%


\begin{minipage}[h!]{0.49\textwidth}
\centerline{\epsfysize 5.7 cm
\epsfbox{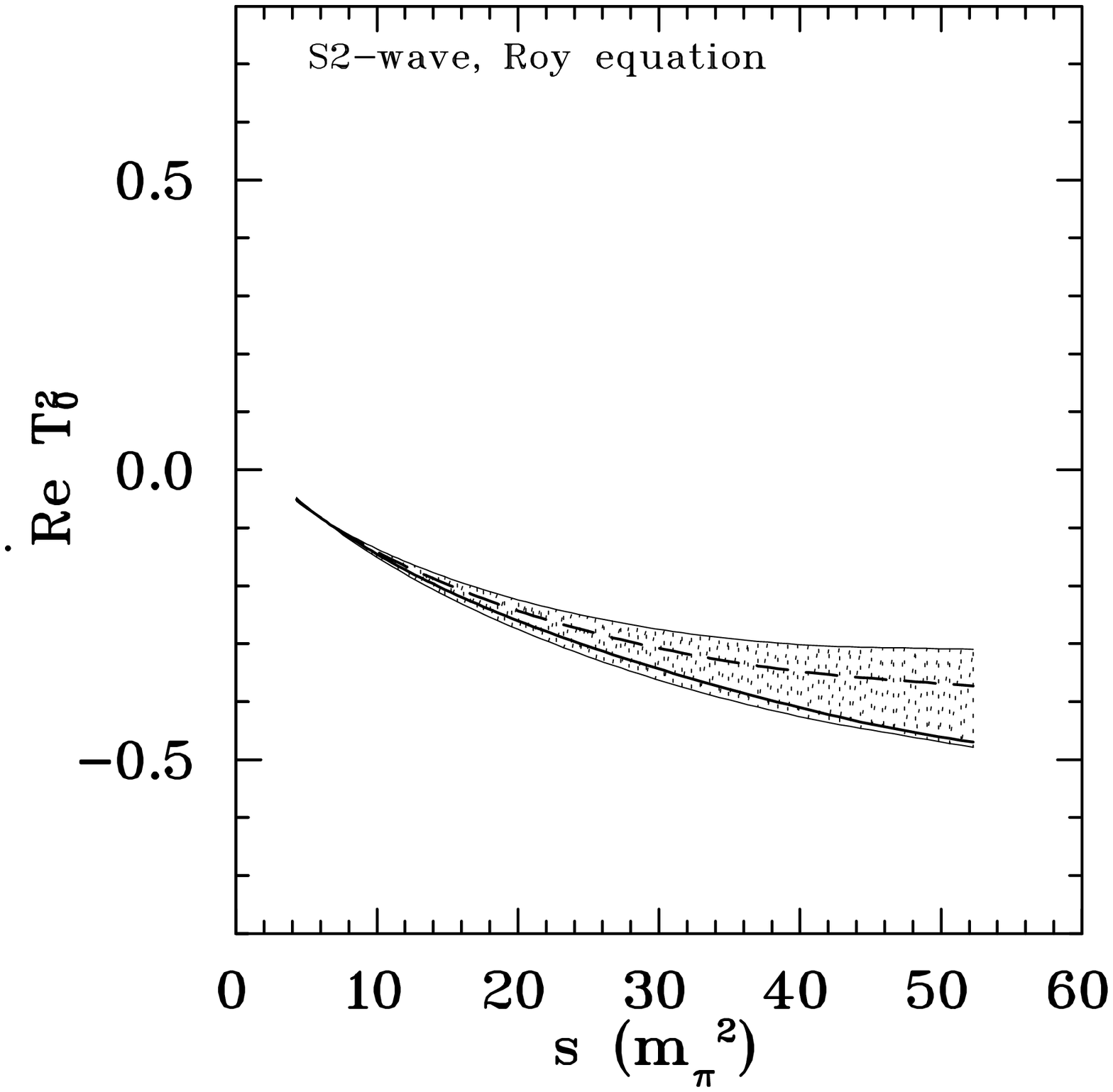}}
\end{minipage}
\begin{minipage}[h!]{0.49\textwidth}
\centerline{\epsfysize 5.7 cm
\epsfbox{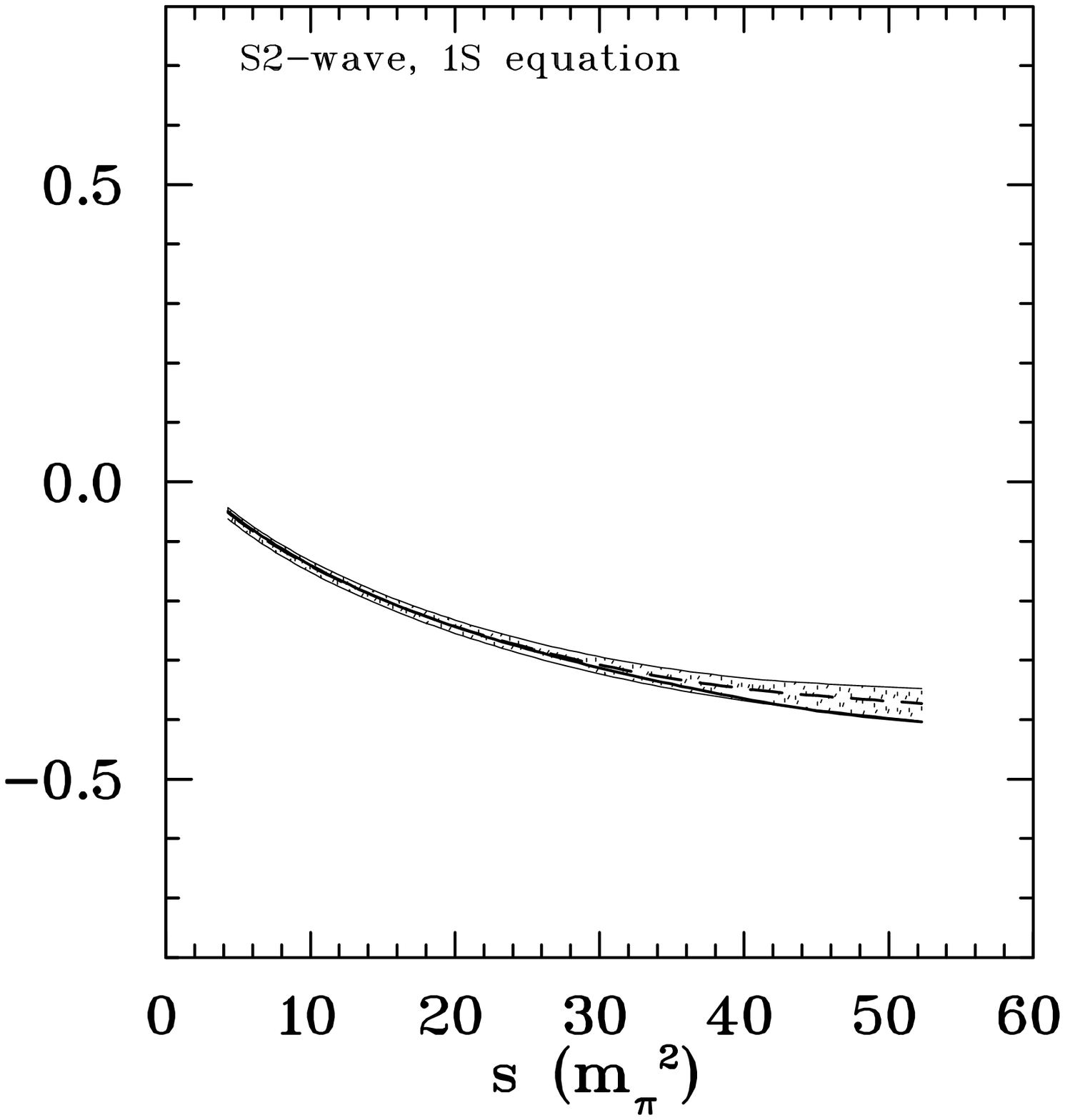}}
\end{minipage}


\caption{\label{RoyPaco} Comparison of results from Roy's and 1S 
equations for waves $S0$, $P$ and $S2$. 
The gray bands correspond to the errors for these equations.
The dashed and solid lines represent the input and output amplitudes, respectively.}
\end{center}
\end{figure}

Figure \ref{RoyPaco} presents a comparison between the total output amplitudes
from Roy's and GKPY equations for the $S0$, $P$ and $S2$ waves.
The error bands plotted around the input amplitudes represent the difference
between the input and the output, and were generated using a
Monte Carlo Gaussian sampling of all parameters in the Constrained
Fit to Data (within 6 standard deviations). The asymmetric errors correspond
to the independent left and right widths of the generated distribution
for $10^5$ events. 
As can be seen on Fig.~\ref{RoyPaco}, even though the CFD set of amplitudes
was not constrained to fulfill the GKPY equations, they are very well
satisfied, with
the differences between input and output amplitudes being generally smaller
than in the case of Roy's equations. A new Constrained Fit to Data in which the
parametrizations are constrained not only to FDR, sum rules and Roy's Eqs.,
but also to the GKPY equations, and in which the functional form of the
parametrizations is refined is in progress.

Especially relevant is that above $s \approx 8 m_\pi^2 \approx (400\mbox{ MeV})^2$
the error bands in all the three waves
for the GKPY equations are significantly narrower than those obtained for Roy's equations.
As already explained, this comes from the fact that the term $ST(s)$
is a constant, and does not grow
with energy, as it was the case with Roy's equations.
The errors for the GKPY equations in the three waves
come almost completely from the $KT(s)$ terms.
As their absolute values are smaller than those of the corresponding
functions for Roy's equations, their errors are also smaller
above $s\approx 8 m_\pi^2$.
Comparing the non-symmetric widths of the error bands for
Roy's equations on Fig.~\ref{RoyPaco} with those 
calculated in~\cite{KJYII} as 
\begin{equation}
\Delta Re t^I_\ell = \sqrt{\sum\limits_{j} \delta_j^2},
\end{equation}
where $\delta_j$ is the error coming from varying the $j$-th parameter
of the CFD set, one obtains, as expected, quite similar results.
This is because the errors coming from each individual parameter are
small, and the number of parameters is large.
In principle the usual Monte Carlo Gaussian sampling keeps
a better detail of the correlations, and since they provide
asymmetric errors. 
they will also be used to estimate our errors.

\section{Conclusions}

We have briefly introduced a new set of
dispersion relations for $\pi\pi$ partial waves, called GKPY for brevity,
with one subtraction and crossing symmetry implemented in a similar way
as it is done in Roy Eqs. Both GKPY and Roy Eqs. provide the amplitude
as a sum of three kinds of contributions:
``subtraction terms'' which contain the subtraction constants, ``kernel terms''
that contain the dispersive integrals of $S0$-, $P$- and $S2$-waves up to
a given energy, and the ``driving terms'' that contain contributions 
from the rest of waves and high energies.
As we have shown here, in the case of the new GKPY, the dominant role
for the $S0$-, $P$- and $S2$-waves is played by
the so called kernel terms. 
They contain information on the energy dependence of other partial waves below 1420 MeV.
In contrast, for standard Roy's equations strong 
cancellations between kernel and subtracting terms occur
in the $S0$ and $S2$ partial waves, since
these terms are several times bigger
than the corresponding ones in the GKPY equations.
Actually, in Roy Eqs. the subtraction terms grow quadratically with energy
and the large experimental uncertainty on the scattering lengths thus propagates
to higher energies as a large source of error. Hence, 
despite Roy Eqs. provide a stringent test for amplitudes at low energy, 
the GKPY provide an even stronger constraint above roughly $s^{1/2} = 400$~MeV,
where they have 
significantly smaller errors than Roy's equations, given the same input.
We have also shown here that, although the dependence on the less known 
high energy input
is less suppressed than in standard Roy Eqs., the driving terms are still
small in comparison with the KT and ST.

In conclusion, we have shown that GKPY are a good tool to constraint 
$\pi\pi$ amplitudes in the intermediate energy region. A full data analysis
using amplitudes constrained to satisfy simultaneously 
Forward Dispersion Relations, Roy and GKPY equations is in progress.
Once the data analysis is completed,
GKPY should also provide a very precise analytic extension to the complex plane
that could be relevant for the study of the poles associated to light resonances.


\section{Acknowledgements}

This work is dedicated to memory of our good friend Prof. F.~J.~Yndurain,
who very unfortunately passed away during the completion of this work.
We thank Spanish research contracts PR27/05-13955-BSCH, FPA2004-02602, UCM-CAM
910309 and BFM2003-00856 for partial financial support.

\vspace{-0.3cm}


\end{document}